\documentclass[aps,twocolumn,floats,prl,nofootinbib]{revtex4}
\usepackage{graphics,graphicx,epsfig}
\usepackage{amssymb,color}
\usepackage{amsmath}
\usepackage{epsf,epstopdf,wrapfig}

\begin{document}

\title{Action at a distance in transcriptional regulation}

\author{William Bialek,$^{a,b}$  Thomas Gregor,$^{a,c}$ and Ga\v{s}per Tka\v{c}ik$^d$}

\affiliation{$^a$Joseph Henry Laboratories of Physics, and Lewis--Sigler Institute for Integrative Genomics, Princeton University, Princeton NJ 08544 USA\\
$^b$Initiative for the Theoretical Sciences, The Graduate Center, City University of New York, 365 Fifth Ave, New York NY 10016\\
$^c$Department of Developmental and Stem Cell Biology UMR3738, Institut Pasteur, 75015 Paris, France\\
$^d$Institute of Science and Technology Austria, Am Campus 1, A-3400 Klosterneuburg, Austria}

\begin{abstract}
There is increasing evidence that protein binding to specific sites along DNA can activate the reading out of genetic information without coming into direct physical contact with the gene.  There also is evidence that these distant but interacting sites are embedded in a liquid droplet of proteins which condenses out of the surrounding solution.  We argue that droplet--mediated interactions can account for crucial features of gene regulation only if the droplet is poised at a non--generic point in its phase diagram.  We explore a minimal model that embodies this idea, show that this model has a natural mechanism for self--tuning, and suggest direct experimental tests.
\end{abstract}

\date{December 18, 2019}

\maketitle

In multicellular organisms, the transcription of genes into messenger RNA is controlled by the binding of transcription factor proteins to ``enhancer'' sites that can be separated from the gene by tens of thousands of base pairs along the DNA sequence \cite{deLaat+al_13,levine+al_14,furlong+al_18,stadhouders+al_19,vanSteensel+al_19,schoenfelder+al_19}.  Close approach of  enhancers to their target promoters has been inferred from cross--linking experiments \cite{misfud+al_15}, and there is direct evidence that the action of the enhancer requires physical proximity to the promoter site where transcription is initiated \cite{chen+al_18}. But proximity is not contact: the most recent measurements indicate that the enhancers and their target promoters remain separated by   $150-350\,{\rm nm}$ even during active transcription \cite{chen+al_18,mateo+al_19,alexander+al_19,heist+al_19,barinov+al_20}.

\begin{figure}[b]
\centerline{\includegraphics[width = 0.85\linewidth]{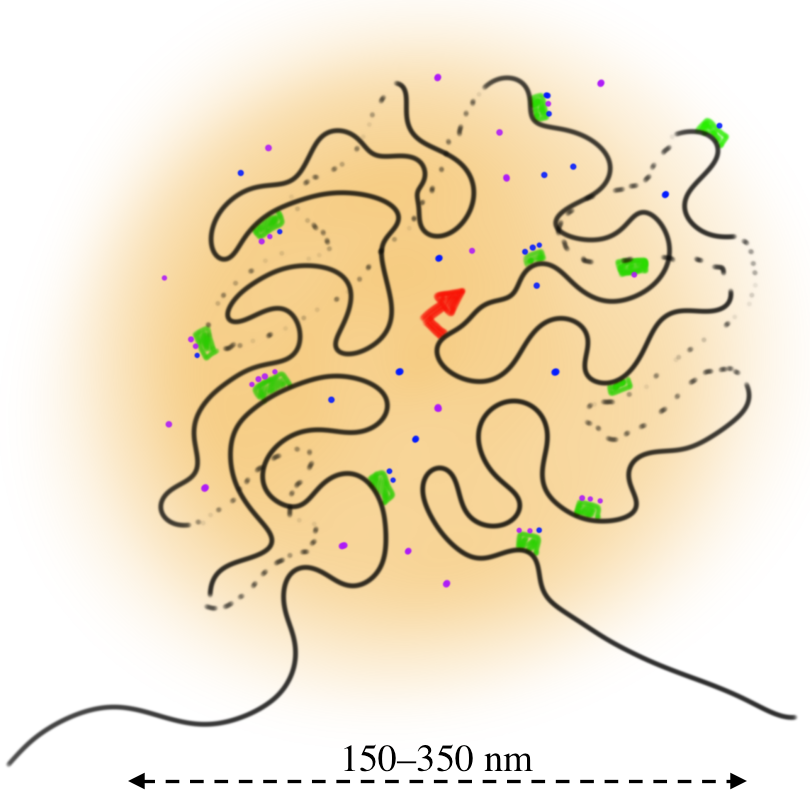}}
\caption{The DNA strand (black) surrounds a condensed droplet (orange).  Promoter site is marked by an arrow (red), enhancer sites as blocks (green) with transcription factors (magenta and blue) both bound to these sites and freely diffusing.
\label{schematic}}
\end{figure}

How is the apparent action at a distance possible? Interactions between the enhancer and promoter could be transmitted along the length of the DNA molecule, but it seems more plausible that this interaction is transmitted across the shorter three dimensional distance \cite{benabdallah+al_15}.    Recent observations indicate that there is a medium for this transmission,   a condensed droplet of the protein ``mediator'' which surrounds the promoter \cite{cho+al_18}; these droplets also contain high concentrations of RNA polymerase   \cite{cisse+al_13}, are associated with foci of active transcription \cite{cho+al_16}, can form in vitro \cite{boija+al_18}, and contain other co--activating factors \cite{sabari+al_18}. We propose that the droplet acts as a larger scale version of an allosteric protein \cite{monod+al_63,monod+al_65,perutz_90}: in the same way that the protein structure allows binding of small molecules at one site to influence binding or enzymatic activity at a distant site,  the droplet would allow binding of transcription factors (TFs) at an enhancer site to influence activity at the distant promoter site (Fig \ref{schematic}).  We will argue that this is possible only if the droplet is at a non--generic point in its phase diagram, and that the collective interactions among the enhancer sites can drive the system toward such points.

Two facts will be crucial to our discussion.  First, to the extent that the mediator droplets are similar to other examples of intracellular phase separation \cite{banani+al_17}, they will be liquid--like \cite{shin+al_18}, and thus in general will not transmit structural changes across hundreds of nanometers. Second, gene expression can be controlled in a quantitative, graded fashion in response to changing concentrations of transcription factors \cite{colman+al_05,takahashi+al_08,giorgetti+al_10,rogers+al_11}.

If we did not have the constraint of graded responses, we could imagine that binding of TFs to enhancer sites triggers droplet condensation, and that this is the essential mechanism of regulation, as  proposed  for ``super--enhancers''  \cite{hinsz+al_17,sabari+al_18}.   But this is an all--or--none mechanism, and it is difficult to harness the triggering of phase separation to generate a quantitatively graded response to changes in TF  concentration. The existence of droplets by itself does not solve the problem.

Even if transcription requires droplet condensation, there are pathways for regulation once the droplet has formed.   In eukaryotes transcription involves a very large number of different proteins, and it is plausible that many of these components condense into the droplet.  With multiple components the phase diagram  is more complicated than just two phases \cite{sear+cuesta_03,mao+al_19}, so  droplets can condense and still have additional degrees of freedom  related to the addition or expulsion of different molecular species.  Let us summarize these variables by an order parameter $\phi({\vec r})$, which can vary with position $\vec r$ inside the droplet.  These are the degrees of freedom that can propagate interactions through the droplet.  

The simplest model envisions a set of $K$ identical binding sites for a single class of transcription factors (at positions $\vec r_{\rm i}$), plus one promoter site (at $\vec r_a$), all embedded in a droplet.    These binding sites typically will be arrayed across multiple enhancers, all of which can contribute to regulating transcription.  The relevant variables are $\sigma_{\rm i} = \pm 1$ for empty and occupied binding sites, and $A = \{0,1\}$ for the inactive and active states of the promoter. All of these variables couple to the order parameter, and it is important that these couplings are spatially local.   The free energy is then
\begin{widetext}
\begin{equation}
{\cal F} [\phi(\vec r );  \{\sigma_{\rm i}\},\,A ]
= {\cal F}_0\left[  \phi(\vec r )\right] 
+ E_0 A 
- {{k_B T}\over 2} \ln (c/c_0) \sum_{{\rm i}=1}^K  \sigma_{\rm i}
+ \sum_{{\rm i}=1}^K g_{\rm i} \sigma_{\rm i}\phi(\vec r_{\rm i} )  
+ g_{\rm a} A \phi(\vec r_a ) 
,
\end{equation}
\end{widetext}
where  $g_{\rm i}$ is the interaction  between the order parameter and binding of TFs to the enhancers, $g_{\rm a}$ is the interaction  between the order parameter and the active vs inactive state of the promoter, $E_0$ is the free energy difference between the two states of the promoter in the absence of TFs, $c$ is the concentration of these factors, and $c_0$ is the ``bare'' binding constant of the TF to the enhancer sites.

Let's assume that, as in conventional models of allostery, the transmission of information can be described as an equilibrium thermodynamic effect \cite{marzen+al_13,einav+al_16,eqm}.  Hence, we define an effective free energy by integrating out the fluctuations of the order parameter, 
\begin{equation}
e^{-F_{\rm eff}(\{\sigma_{\rm i}\},\,A)/k_B T} = \int {\cal D}\phi \exp\left( - {{\cal F} [\phi(\vec x );  \{\sigma_{\rm i}\},\,A ]\over{k_B T}}\right) ,
\end{equation}
where ${\cal D}\phi$ is the measure for integration over  $\phi(\vec r )$.
$F_{\rm eff}$ is composed of independent and interacting parts, $F_{\rm eff}  = F_0 +  F_{\rm int}$, and to leading order in the couplings we have
\begin{eqnarray}
F_0 &=&   E_0 A
- {{k_B T}\over 2} \ln (c/c_0) \sum_{{\rm i}=1}^K  \sigma_{\rm i}
\label{F0_eqn}\\
F_{\rm int} &=& 
-   \sum_{{\rm i},\,{\rm j} =1}^K {{g_{\rm i} g_{\rm j}}\over{2k_B T}}  C(r_{\rm ij}) \sigma_{\rm i} \sigma_{\rm j} 
- \sum_{{\rm i}=1}^K {{g_{\rm a} g_{\rm i}}\over{k_B T}}  C(r_{{\rm i}a})  \sigma_{\rm i} A 
,\nonumber\\
&&
\end{eqnarray}
where  we choose coordinates so that the average order parameter is zero  in the Boltzmann distribution defined by ${\cal F}_0$, and $C(r)$ is the correlation function of the order parameter fluctuations in this distribution, 
\begin{equation}
\langle \phi(\vec r_{\rm i} ) \phi(\vec r_{\rm j} ) \rangle \equiv C(r_{\rm ij}) ,
\end{equation}
with $r_{\rm ij} = | {\vec r}_{\rm i} - {\vec r}_{\rm j}|$. The question of whether the droplet  transmits information from the enhancer to the promoter becomes the question of whether fluctuations in the order parameter  are correlated over these long distances \cite{sethna_06}.

In liquids at generic parameter values, density fluctuations have a short correlation length $\xi$, so that $C(r) \simeq e^{-r/\xi}$, with  $\xi$ on the nanometer scale, and thus these modes cannot support action at a distance.  There can be additional degrees of freedom associated with the orientational ordering of molecules in the droplet, or with conformational changes of these molecules, but again with generic parameters we expect to find small $\xi$.   The alternative is that the parameters describing the droplet are at a non--generic point in the phase diagram, where the correlation length can become long, and this is the ``critical droplet'' scenario we explore here.

Close to a first--order phase transition the free energy ${\cal F}_0$ has two nearly degenerate minima \cite{sethna_06}.  In a sufficiently small droplet, fluctuations in the order parameter are  dominated by flickering between these minima, and there is an effective surface tension that keeps the entire droplet in one minimum, so that $C(r)$ becomes only weakly dependent on distance \cite{1stOrder}.  Close to a second--order phase transition  the correlation length  diverges, and  $C(r)$ decays very slowly, as power of distance.  Either of these scenarios seems to require some tuning of the droplet parameters, to which we return below.

\begin{figure}[t]
\centerline{\includegraphics[width = 0.8\linewidth]{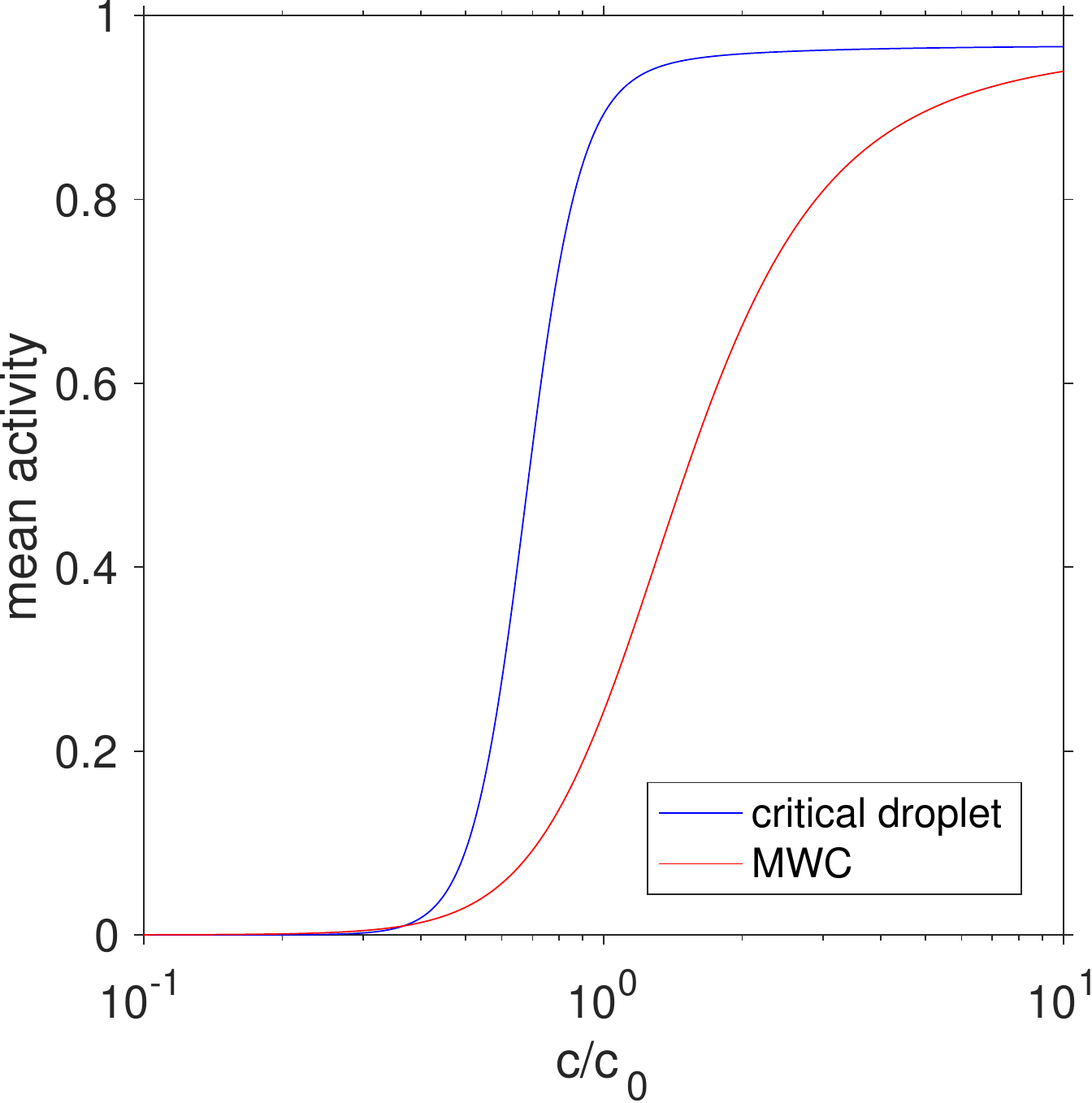}}
\caption{Mean promoter activity as function of the TF concentration.  Results from Eqs (\ref{F0_eqn}) and (\ref{Feff}) with $K=8$ sites,  $J=0.2 k_B T$, $J_a = k_B T$, and $E_0 =k_B T \ln(100)$, compared with the corresponding MWC model ($J_a=0.2 k_B T$, $J=0$).  Note that interaction energies are on the order of $k_B T$ or less, but the droplet generates a very steep response to changing TF concentrations.
\label{meanA}}
\end{figure}

If all the transcription factor binding sites couple to the droplet in the same way, then we should have all the $g_{\rm i} = g$.  We can capture the essential predictions of this model if all the distances $r_{\rm ij}$ and $r_a$ are roughly equal to a typical $R$, in which case we can simplify to
\begin{equation}
F_{\rm int} =
- {J\over 2} \sum_{{\rm i},\,{\rm j} =1}^K \sigma_{\rm i} \sigma_{\rm j}  -  J_a \sum_{{\rm i}=1}^K \sigma_{\rm i} A  ,
\label{Feff}
\end{equation}
with two parameters $J =  ({{g^2 }/{k_B T}})  C(R)$ and $J_a = ({{g g_{\rm a}}/{k_B T}} ) C(R)$. To gain intuition, we note that if $J=0$ then Eq (\ref{Feff}) becomes identical to the Monod--Wyman--Changeux (MWC) model for allosteric proteins \cite{monod+al_65,marzen+al_13}, with the $A = 1/0$ switch in promoter activation playing the role of the R/T conformational transition \cite{monod+al_65,perutz_90}.   A critical droplet thus generates cooperative interactions among distant sites that are similar, qualitatively, to more familiar examples of allostery (Fig \ref{meanA}).

Transcriptional activators become repressors just by changing the sign of the coupling $g_{\rm i}$, which describes interaction of the TF with the surrounding droplet, and  is determined by the face of the protein opposite from the DNA binding domain.  A generalization is to imagine that we have $K_a$ binding sites for activators at concentration $c_a$ and $K_r$ sites for repressors at concentration $c_r$.  Then at low repressor concentrations there is cooperative activation, but at higher repressor concentration the system approximates a switch that depends on the ratio of powers of the concentrations, $c_a^{K_a}/c_r^{K_r}$.

A second generalization is to have multiple nearby binding sites within one enhancer interact more strongly,  perhaps through additional degrees of freedom, and then let the emergent states of multiple enhancers couple  to the droplet.  The system could then approximate logical operations corresponding to combinations of ANDs within each enhancer and ORs among enhancers.

An important implication of this model is that transcription is regulated not by the binding of transcription factors to individual binding sites, but rather by an integrated signal from multiple binding sites that are distributed around the droplet of size $R$.  If we think of the transcriptional output as a ``measurement'' of the TF concentration, then  the accuracy of this measurement is limited by the random arrival of molecules   \cite{berg+purcell_77,bialek+setayeshgar_05,vanzon+al_06};  the smallest concentration differences $\delta c$ to which a system can respond reliably is given by
\begin{equation}
{{\delta c}\over c} \simeq {1\over{\sqrt{D \ell c \tau}}} .
\label{bplimit}
\end{equation}
where $c$ is the background concentration, $D$ is the diffusion constant, $\tau$ is the time over which the system can average, and $\ell$ is the linear size of the sensitive element.   If the response is driven by a single binding site, then $\ell$ is the size of that site, but if the system integrates over many binding sites, then $\ell$ can approach the linear dimensions of the entire array of sites, in our case the size of the droplet, which is ${\sim} 100\times$ larger than individual binding sites.   From Eq (\ref{bplimit}), responses which would require hours of integration at a single site thus become reliable in minutes.  Transcription factor concentrations are so low that this difference can be crucial \cite{bialek+setayeshgar_05,gregor+al_07}.

\begin{figure}
\centerline{\includegraphics[width = 0.8\linewidth]{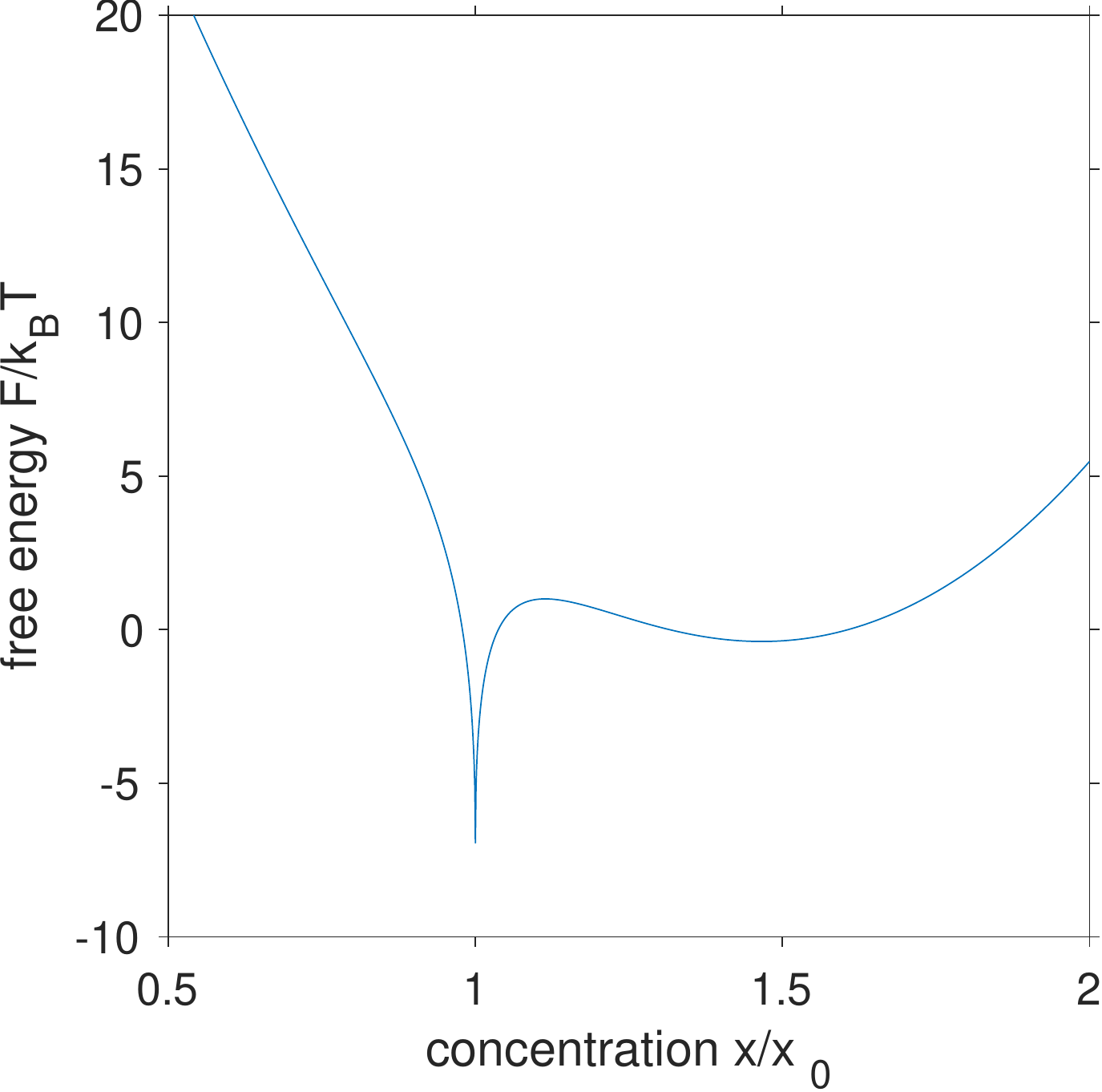}}
\caption{Free energy as a function of concentration in the droplet, from Eq (\ref{FvC_eqn}), with $\bar n = 100$, $\Delta x_0  = x_0/2$,  $A = 0.25$, $R/a = 5$, and $\nu= 1/2$. Note the weak minimum at the $x = x_0 + \Delta x_0$,  set by the chemical potential, which is dominated by the minimum at criticality, $x = x_0$.
\label{FvsC}}
\end{figure}

Poising a condensed droplet near a critical point seems to require fine tuning of its parameters.  Cells can exert exquisite control over protein and nucleic acid concentrations \cite{gregor+al_07,petkova+al_14}, but matching the concentrations of crucial molecules to their critical values still seems difficult.   In our case, however, there is a thermodynamic driving force that pushes the system toward conditions where correlation lengths are long.   To estimate this effect, let's  assume that the droplet has a critical point when one of its components is at concentration $x_0$.  The chemical potential of the surrounding solution holds the concentration close to some mean concentration $\bar x$, and variations $\Delta x$ around this mean cost a free energy $F_1 \simeq (\bar n k_B T /2)  (\Delta x / {\bar x})^2$, where $\bar n \simeq R^3 \bar x$ is the mean number of molecules in the droplet.  But at a concentration $x$ the correlation length will be 
$\xi = a | {{x_0}/({x - x_0}) }{|}^\nu$ \cite{sethna_06}, where $a$ is a microscopic length scale.
Away from criticality, the gain in free energy from interaction among $K$ binding sites is $F_2 \simeq -(J_0/2) K(K-1)e^{-R/\xi}$, and in total we have
\begin{equation}
{{2F}\over {\bar n k_B T}}   \simeq  (\Delta x / {\bar x})^2 - A \exp\left[ - {R\over a} {\bigg |} {{x_0}\over{ \Delta x_0 + \Delta x}}{\bigg |} ^\nu
\right] ,
\label{FvC_eqn}
\end{equation}
with $\Delta x_0 = \bar x  - x_0$  and $A = ({{J_0}/{k_B T}}) {{K(K-1)}/{\bar n}}$. The dominant component of the droplet is present at only $\bar n \sim 100$ \cite{cho+al_18}, so with $K \sim 10$ binding sites it is easy to have $A \sim 1$; to be conservative we consider $A = 0.25$.  We could plausibly have $R\sim 150\,{\rm nm}$ and the molecular scale $a\sim 5\,{\rm nm}$, but again to be conservative we choose  $R/a = 5$.  Assuming that the chemical potential alone sets $\bar x = 1.5 x_0$, we see that the possibility of mediating interaction among binding sites creates a sharp minimum of the free energy at the critical point, sufficient to pull the system very close to criticality (Fig \ref{FvsC}).

Taking this thermodynamic driving force seriously, we note that when transcription is active, enhancer binding sites with states that are ``aligned'' to this activation have a free energy that is lower  by $J_{\rm a} \propto C(R)$, and since $C(R)$ decreases with $r$ this generates a small force pulling the enhancer toward the promoter.  In contrast, enhancers in states that are not contributing to activation of transcription have a free energy that is higher by $J_{\rm a}$ and a force pushing enhancer and promoter apart.  These small forces are balanced by a stiffness, which also determines the thermal fluctuations in the enhancer--promoter distance.  The result is that aligned vs anti--aligned enhancers should be at different mean distances from the promoter site, and this displacement is  $\Delta R /R\sim (J_{\rm a}/k_B T)(\delta R /R)^2 $, where $\delta R$ is the standard deviation of the distance $R$, and we assume that $d\ln C(R)/d\ln R \sim 1$.  These displacements should be directly observable, for example by measuring the positions of   different enhancers for the pair-rule genes in the early fly embryo \cite{stripes}.  More generally this suggests that single--molecule observations of enhancer motions could be connected, quantitatively, to the energetics of cooperative transcriptional activation.

To summarize, a large number of transcription factor binding sites, embedded in a droplet that surrounds the promoter, will generate cooperative regulation if the droplet is poised near special points or lines in its phase diagram where correlation lengths become long.  
In this scenario the droplet functions much like an allosteric protein, but this is possible only because of the proximity to criticality.  This is similar to the long--ranged interactions between proteins that we expect to see in a membrane \cite{machta+al_12} if lipid compositions are close to a critical point, as observed \cite{veatch+al_08,honerkamp-smith+al_09,rayermann+al_17}; it has also been suggested that chromatin itself is close to a sol/gel phase boundary \cite{khanna+al_18}.  There is a much wider range of ideas about how criticality could play a role in biological function \cite{mora+bialek_10,munoz_18}, but what is special in our example is that we have identified, as an intrinsic part of the functional behavior, a mechanism that drives the system toward its critical point, and perhaps this is more general.  Consequences of this thermodynamic driving force should be directly observable in the physical positions of enhancer and promoter sites.

\begin{acknowledgments}
We thank L Barinov, SA Kivelson, and MS Levine for helpful discussions.  This work was supported in part by the US National Science Foundation, through the Center for the Physics of Biological Function (PHY--1734030) and Grant PHY--1607612; by National Institutes of Health Grants P50GM071508, R01GM077599, and R01GM097275; and by Austrian Science Fund grant FWF P28844.
\end{acknowledgments}


\begin{thebibliography}{}
\bibitem{eqm}
Many allosteric proteins are   enzymes, and  the non--equilibrium nature of an active catalyst could make an equilibrium description impossible.  In fact the equilibrium description seems to work, so we start here.
\bibitem{1stOrder}
There is a length scale formed from the ratio of this surface tension to the free energy difference per unit volume between the two minima, which vanishes on the line of first--order transitions, so the length scale diverges.  A problem in this scenario is to be sure that flickering between the minima is fast enough.
\end{thebibliography}
\end{document}